\documentstyle[aps,pre]{revtex}
\begin{document}
\draft
\title{ Roughness-induced wetting
}
\author{
Roland R. Netz\cite{address} and David Andelman
}

\address{
School of Physics and Astronomy, 
Raymond and Beverly Sackler Faculty of Exact Sciences,\\
Tel Aviv University,
Ramat Aviv 69978, Tel Aviv,  Israel}
\date{\today}
\maketitle

\begin{abstract}
We investigate theoretically the possibility of a
wetting transition induced by geometric roughness of a solid substrate
for the case where the flat substrate does not show a wetting layer.
Our approach makes use of a novel closed-form expression which relates
the interaction between two sinusoidally modulated  interfaces 
to the interaction between two flat interfaces.
Within the harmonic approximation, we find that roughness-induced 
wetting is indeed possible if the substrate roughness,
quantified by the substrate surface area, exceeds a certain threshold. 
In addition, the molecular
interactions  between the substrate and the wetting substance  have 
to satisfy several conditions. These results are expressed
in terms of a lower bound  on the wetting potential for a flat 
substrate in order for roughness-induced wetting to occur.
This lower bound has the following properties:
A minimum is present at zero or very small separation 
between the two interfaces, as characteristic for the non-wetting
situation in the flat case. Most importantly, the
wetting potential needs to have a pronounced
maximum at a separation comparable to the amplitude of the
substrate roughness. These findings are in agreement
with the experimental observation of 
roughness-induced surface premelting at  a glass-ice interface
as well as the calculation of the dispersion interaction for
the corresponding glass-water-ice system.
\end{abstract}
%\narrowtext

\section{Introduction}

The phenomenon of wetting has been the subject of intense
attention and a fairly good understanding of the basic concepts 
and mechanisms has emerged\cite{deGennes,Dietrich,Schick}.
In the simplest case, a solid inert surface is put in contact
with an under-saturated vapor of a second substance.
Typically, a molecularly thick liquid-like film 
will form on the substrate surface due to favorable molecular
interactions.
The liquid film is in equilibrium with its
under-saturated vapor, thus giving rise to a second interface
(the emerging liquid-vapor interface), 
referred to hereafter as the liquid interface. 
Depending on the detailed molecular interactions between all three 
phases and the resulting interfacial energies, the liquid film can either
grow to macroscopic thickness  or remain finite as coexistence between 
the liquid and its vapor is approached. The first situation
corresponds to {\em complete wetting}  with a diverging film thickness,
the second case is called {\em incomplete} or {\em partial wetting}.

Two other, closely related situations are possible:
(i) The liquid layer (e.g., water) on the inert substrate  
can be in equilibrium with its solid phase (ice)  at temperatures
below the melting point. In this case
the vapor is replaced by a solid and the appearance  of a 
thin liquid layer between the  substrate and the solid phase
indicates  interfacial premelting. Note that the third  phase (the ice)
is entirely different from the solid substrate.
(ii) The substrate itself can be a solid in equilibrium
with its vapor phase. Here the formation of a thin liquid layer as
three-phase coexistence is approached corresponds to surface
premelting\cite{Dash}.
The phenomenological description  of these scenarios
does not  differ from the wetting
situation, and one finds the analogous phenomena corresponding to
partial and complete wetting.

In early theoretical studies, the solid substrate was assumed
to be flat and homogeneous. However, in most experimental and technological
situations the substrate is both rough and inhomogeneous.
For complete wetting upon approaching  coexistence, 
where the liquid forms a thin
and continuous film, the influence of substrate roughness  and chemical
disorder has recently been investigated  
theoretically\cite{Andelman,Pfeiffer,Kardar,Napiorkowski,Korocinski}
and experimentally\cite{Beaglehole,Garoff,Tidswell,Muller} in great detail.
It was found that heterogeneity and roughness of the solid substrate in
conjunction with long-range
van der Waals interactions cause equilibrium undulations of the
liquid film surface. Surface tension, on the other hand, acts as a
damping mechanism which reduces the amplitude of undulations for thicker films.
The theoretical results\cite{Andelman,Napiorkowski}
were verified recently in small-angle x-ray 
scattering\cite{Garoff,Tidswell,Muller}.

Yet another realization of the wetting phenomenon is obtained
if a {\em non-volatile} liquid is spread on a solid surface;
in technological applications, the liquid might be a paint or a lubricant.
In this case, the liquid is neither  in phase equilibrium 
with the solid substrate nor with the gaseous phase, and the total
amount of liquid on the substrate is a conserved quantity\cite{deGennes}.
In the complete wetting situation, the liquid forms a continuous film
on the substrate; in the partial or incomplete wetting
case, the liquid forms droplets with the contact angle being determined
by the interfacial energies between the three phases meeting
at the contact line\cite{Indekeu}. Roughness of the substrate has been shown
to cause contact angle hysteresis for advancing and receding
contact lines\cite{deGennes,Leger}.

A lot of work was specifically concerned with the interfacial
and surface premelting properties of ice, due to its atmospheric
and environmental consequences\cite{Dash2}.
Surface premelting of ice has been observed by a variety 
of experimental techniques\cite{Beaglehole3,Furukawa,Lied}, 
and it is now believed that
complete surface premelting (i.e., macroscopic
growth of the surface liquid layer as the melting
temperature is approached) only occurs for some orientations
of the crystal surface and only if the vapor phase is diluted 
with air\cite{Elbaum}.
Interfacial premelting of ice, giving rise to the
low sliding friction of ice, has been deduced from
wire regelation at low temperature\cite{Gilpin}, and
from viscosity measurements between surfaces of
ice and quartz\cite{Barer}; it also forms the basis
for frost heave in frozen soils\cite{Wilen2}.
Complete interfacial premelting between ice crystals
and a glass substrate was detected by
ellipsometry\cite{Furukawa2,Beaglehole2}.
The geometric structure of the glass surface was shown
to play  a vital role in this premelting phenomenon;
in a series of experiments, the surface has been roughened by
exposition to fluoric acid for different amounts of time,
leading to surfaces with varying characteristic 
height-fluctuation amplitudes and wavelengths\cite{Beaglehole2}.
For flat glass substrates the premelting was shown
to be incomplete, while complete premelting was exhibited
for glass substrates with a threshold amount of 
micro roughness\cite{Beaglehole2}.

The latter experimental observation
motivated us to explore theoretically
the possibility of a 
{\em roughness-induced} complete wetting or premelting transition. 
This describes the  situation in
which the {\em flat} substrate, for a given temperature,
is not covered with a 
macroscopic liquid layer as coexistence is approached
(corresponding to partial wetting),
but, at the same temperature, is completely wet
if the roughness of the substrate exceeds a certain threshold
(in what follows we will use the wetting terminology both
for the phenomena of premelting and wetting).
In this paper we critically examine the conditions under
which such a phenomenon can occur. 
As a result, roughness-induced wetting is indeed possible if the involved
materials have the following properties: i) The tension of the substrate-vapor
interface has to be larger than that of the substrate-liquid interface.
ii) The surface area increase of the solid substrate due to its roughness
has to exceed a certain threshold, which depends on the
interfacial tensions of all three phases\cite{Beaglehole2}.
iii) The effective interaction between the two interfaces bounding the
liquid layer for the flat case has to have a maximum
for separations of about the amplitude of the substrate roughness.
We also find that roughness-induced wetting is most likely to occur
when the substrate roughness just exceeds a certain threshold value and will
disappear for  very large amplitudes of the roughness.

The outline of this paper is as follows: 
In Sec. II we introduce the model and review some nomenclature for the
case of a flat substrate.
In Sec. III we extend the analysis to the case of a rough substrate.
We first give necessary and rather general conditions 
under which roughness-induced  wetting is possible.  Using a formula
which describes the van der Waals interaction between two sinusoidal
surfaces, we then construct a sufficient condition
for roughness-induced wetting  in the form
of a lower bound for the interaction between two flat interfaces.
Finally, Sec. IV contains the discussion. 

\section{Flat Substrate}
Consider the situation as illustrated in Fig. 1, where a thin
liquid film  intrudes in between an inert  solid surface and
a top phase. The top phase  can be either a vapor or a solid in thermodynamic 
equilibrium with the liquid film.
The solid substrate, on the other hand, is completely frozen and far 
from its melting point. In this section, we will review results for 
an ideal solid substrate; namely, molecularly flat and
homogeneous. Using the convention of labeling all physical quantities 
with a {\em zero} subscript for the  flat case,
the free energy per unit area can be written as
\begin{equation}
{\cal F}^0 (\ell)=
\gamma_{SL} + \gamma + P^0(\ell) + \mu \ell
\end{equation}
where $\gamma_{SL}$ denotes the 
solid-liquid interfacial tension, and $\gamma$ denotes
the interfacial tension between the liquid and the top phase.
The parameter  $\mu$ is the chemical potential difference between
the liquid and top phases. Alternatively, it 
could also correspond to a Lagrange
multiplier  controlling the film thickness for non-volatile
liquids with conserved total volume.
The potential $P^0(\ell)$ represents the interaction per unit area
between the two flat interfaces with a separation of $\ell$
and can be viewed as a thickness-dependent correction to the 
interfacial energies, depending both on the short and long-ranged
parts of the molecular interactions\cite{Dietrich2}.

In the simplest approach,
assuming pair-wise additive interactions between 
molecules and uniform densities in all coexisting phases, 
$P^0(\ell)$ can be 
expressed as 
\begin{equation}
P^{0}(\ell) \equiv
\int_{\ell}^{\infty} \mbox{d}z
\int \mbox{d}^{2}\mbox{\boldmath $\rho$}
\int_{-\infty}^{0} \mbox{d}z' \;\;
w(\mbox{\boldmath $\rho$},z-z')
\end{equation}
where \(w(\mbox{\boldmath $\rho$},z)\) corresponds
to the local interaction energy difference per unit volume
squared  between the solid and the 
third phase. Dropping some constant terms, it  can be written as
\begin{equation}
w({\bf r})=
n_L^2 U_{LL}({\bf r})-
n_L n_S  U_{LS}({\bf r})-
n_L n_T  U_{LT}({\bf r})+
n_T n_S  U_{TS}({\bf r})
\end{equation}
Here, $U_{ij}({\bf r})$ are the pair interactions between 
molecules 
and the $n_i$ are the particle number densities for each phase,
where $i$ and $j$ are any of the relevant
phases: solid (S), liquid (L), top phase (T)\cite{Andelman}.
In a more realistic approach, one calculates (2) directly using the 
Lifshitz continuum theory of dispersion interactions\cite{foot1}.

In general, $P^0(\ell)$ is expected to vanish for $\ell \rightarrow \infty$
as the two interfaces become decoupled
and approaches a finite value for  $\ell \rightarrow 0$\cite{foot2}.
One therefore defines
\begin{equation}
P^0(\ell)  = \left\{ \begin{array}{lll}
        0 & \mbox{for }     & \ell \rightarrow  \infty  \\
        S    & \mbox{for }  & \ell \rightarrow  0 \\
                \end{array} \right.
\end{equation}
where $S$ is traditionally called the {\em spreading coefficient} and
is given by
\begin{equation}
S \equiv \gamma_{ST}-\gamma_{SL} - \gamma
\end{equation}

\noindent
The above definition leads to ${\cal F}^0(0)=\gamma_{ST}$, just
as one would expect: in the absence of any liquid, the total free
energy is given by the interfacial tension between the solid and the
top phase\cite{foot3}. On the other hand, for 
an infinite layer of liquid coexisting with the top phase,  
one finds ${\cal F}^0(\infty)=\gamma_{SL}+\gamma$ for $\mu=0$,
i.e., the two interfaces bounding the liquid
do not interact and the total free energy is given by the sum of the two
interfacial energies alone.
One can notice that positive values of $S= P^0(0)-P^0(\infty)=
{\cal F}^0(0)-{\cal F}^0(\infty)$ correspond to a situation where
an  infinite  liquid layer is energetically preferred over 
a vanishing liquid layer. Indeed, neglecting the 
possibility of additional minima of 
$P^0(\ell)$ at intermediate values of $\ell$, positive and negative
values of $S$ correspond to wetting and non-wetting cases, respectively.
On the other hand, 
in the case of a non-vanishing chemical potential $\mu$, the minimum of the
free energy will always be at finite film thickness, even for 
positive spreading coefficient$S$\cite{Brochard}.

In the following, we will be exclusively concerned with the non-wetting case,
i.e., $S<0$. 
It will be convenient to modify the definition of the free energy 
slightly and to take the infinitely thick liquid layer  as the reference state.
The free energy difference, defined by
$\Delta {\cal F}^0(\ell)\equiv {\cal F}^0(\ell)- {\cal F}^0(\infty)$
and in the case of vanishing chemical potential, $\mu=0$,
is given by
\begin{equation}
\Delta {\cal F}^0(\ell)=
P^0(\ell)
\end{equation}
with the limiting values
\begin{equation}
\Delta {\cal F}^0 (\ell)  = \left\{ \begin{array}{lll}
        0  & \mbox{for }     & \ell \rightarrow \infty  \\
        S    & \mbox{for }  & \ell \rightarrow  0 \\
                \end{array} \right.
\end{equation}
Clearly, a wetting situation is realized if $\Delta {\cal F}^0 (\ell) >0$
holds for all $\ell < \infty$.

\section{Rough Substrate}

We introduce now the necessary framework to describe
wetting on geometrically rough solids\cite{foot4}. 
The free energy per unit projected area 
for a liquid film on a rough solid substrate (see Fig. 2)
can be written as 
\begin{equation}
{\cal F}(\mbox{\boldmath $\rho$},[\zeta_{L}])=
\sqrt{1+[\nabla\zeta_{S}(\mbox{\boldmath $\rho$})]^2} \; \gamma_{SL}+
\sqrt{1+[\nabla\zeta_{L}(\mbox{\boldmath $\rho$})]^2} \; \gamma+
P(\mbox{\boldmath $\rho$},[\zeta_{L}])+
\mu[\zeta_{L}(\mbox{\boldmath $\rho$})-
\zeta_{S}(\mbox{\boldmath $\rho$})]
\end{equation}
The solid and liquid surfaces  are parameterized by
$\zeta_{S}(\mbox{\boldmath $\rho$})$ and $\zeta_{L}(\mbox{\boldmath $\rho$})$,
respectively, where \(\mbox{\boldmath $\rho$}\) is a two-dimensional
vector in a reference plane.
The interaction term $P(\mbox{\boldmath $\rho$},[\zeta_{L}])$ 
is a generalization of $P^0(\ell)$, Eq. (2), and is 
defined by
\begin{equation}
P(\mbox{\boldmath $\rho$},[\zeta_{L}])=
\int_{\zeta_{L}(\mbox{\boldmath $\rho$})}^{\infty} \mbox{d}z
\int \mbox{d}^{2}\mbox{\boldmath $\rho'$}
\int_{-\infty}^{\zeta_{S}(\mbox{\boldmath $\rho +\rho'$})} \mbox{d}z'
\; \; w(\mbox{\boldmath $\rho'$},z-z')
\end{equation}
Note that expression (9) is a local function of the liquid profile
$\zeta_{L}(\mbox{\boldmath $\rho$})$, 
but a non-local functional of the rough solid 
surface $\zeta_{S}(\mbox{\boldmath $\rho$})$.
For the discussion of wetting behavior it is useful to
average over the in-plane coordinate $\mbox{\boldmath $\rho$}$,
by which we obtain the effective wetting free energy
\begin{equation}
\overline{{\cal F}}(\ell,[\zeta_{L}]) \equiv
\langle {\cal F}(\mbox{\boldmath $\rho$},[\zeta_{L}])
\rangle_{\mbox{\boldmath $\rho$}}
\end{equation}
where we explicitly pulled out the dependence on the average film thickness
\begin{equation}
 \ell \equiv \langle \zeta_{L}(\mbox{\boldmath $\rho$})-
\zeta_{S}(\mbox{\boldmath $\rho$})\rangle_{\mbox{\boldmath $\rho$}}
\end{equation}
This parameter measures the average distance between the two interfaces.
The effective free energy (10) can be written as
\begin{equation}
\overline{\cal F}(\ell,[\zeta_{L}])=
\alpha_S \gamma_{SL} + \alpha_L(\ell,[\zeta_{L}]) \gamma + 
\overline{P}(\ell,[\zeta_{L}]) + \mu \ell
\end{equation}
In analogy to (10),
the effective wetting potential $\overline{P}(\ell,[\zeta_{L}])$ 
is obtained from (9) by averaging over the in-plane coordinate
\begin{equation}
\overline{P}(\ell,[\zeta_{L}])  \equiv
\langle P(\mbox{\boldmath $\rho$},[\zeta_{L}])
\rangle_{\mbox{\boldmath $\rho$}}
\end{equation}
The parameters $\alpha_{S}$ and  $\alpha_{L}(\ell,[\zeta_{L}])$
measure the ratios
between the actual and projected areas of the substrate
surface and liquid interface, respectively, and are defined by
\begin{equation}
\alpha_{S} \equiv \langle \;
\sqrt{1+[\nabla\zeta_{S}(\mbox{\boldmath $\rho$})]^2}
\;\; \rangle_{\mbox{\boldmath $\rho$}}
\end{equation}
\begin{equation}
\alpha_{L}(\ell,[\zeta_{L}])\equiv \langle \;
\sqrt{1+[\nabla\zeta_{L}(\mbox{\boldmath $\rho$})]^2}
\;\;\rangle_{\mbox{\boldmath $\rho$}}
\end{equation}
On the mean-field level considered in this paper, one can take the liquid
interface to assume a fixed profile $\zeta_{L}^*(\mbox{\boldmath $\rho$})$ 
such as to minimize the free energy expression (12).
By construction of the functional (12), this amounts to a
constrained minimization of the free energy 
for a fixed average film thickness $\ell$. 
This yields the minimized free energy, denoted by 
$\overline{\cal F}^*$, as a function of $\ell$,
\begin{equation}
\overline{\cal F}^*(\ell) \equiv 
\min_{[\zeta_{L}]}
\overline{\cal F}(\ell,[\zeta_{L}])=
\overline{\cal F}(\ell,[\zeta_{L}^*])
\end{equation}
The area ratio of the optimal liquid interface
$\zeta_{L}^*$  has the limiting values
\begin{equation}
\alpha^*_{L}(\ell) \equiv \alpha_{L}(\ell,[\zeta_{L}^*])
  = \left\{ \begin{array}{lll}
        \alpha_{S}& 
          \;\;\;\;\;  \mbox{for } & \ell \rightarrow  0  \\
        1    & \;\;\;\;\; \mbox{for }  & \ell \rightarrow  \infty \\
                \end{array} \right.
\end{equation}
since a very thin liquid layer follows the solid substrate roughness
completely whereas a thick enough layer will be essentially flat
(neglecting thermal capillary roughness).
Just as for the flat case, $\overline{P}^*(\ell)\equiv
\overline{P}(\ell,[\zeta_{L}^*])$ 
is expected to vanish for
infinitely separated interfaces, i.e., $\overline{P}^*(\infty)=0$.
The interaction at contact (for $\ell=0$),  is to a first approximation
given by the interaction of the flat case times the surface area ratio
of the rough solid surface, i.e., 
$\overline{P}^*(0)\approx  \alpha_{S} P^0(0)=\alpha_S S$\cite{foot5}.
Defining the free energy difference by
$\Delta \overline{\cal F}(\ell,[\zeta_{L}])\equiv 
\overline{\cal F}(\ell,[\zeta_{L}])- \overline{\cal F}^*(\infty)$,
for which we set  $\mu=0$ (the introduction of a non-zero 
chemical potential is straightforward and will be treated 
separately in Sec. III.C), one finds
\begin{equation}
\Delta \overline{\cal F}(\ell,[\zeta_{L}])=
(\alpha_{L}(\ell,[\zeta_{L}])-1)  \gamma
+\overline{P}(\ell,[\zeta_{L}])
\end{equation}
The limiting values of the free energy $\Delta \overline{\cal F}^* (\ell)$
(obtained by minimizing with respect to the liquid interface profile
$\zeta_{L}$) are given by
\begin{equation}
\Delta \overline{\cal F}^* (\ell)  = \left\{ \begin{array}{lll}
        \alpha_{S}(\gamma_{ST}-\gamma_{SL})-\gamma & 
          \;\;\;\;\;  \mbox{for } & \ell \rightarrow  0  \\
        0    & \;\;\;\;\; \mbox{for }  & \ell \rightarrow  \infty \\
                \end{array} \right.
\end{equation}
It is instructive to define the  effective spreading coefficient, 
$S_{\rm eff} \equiv \Delta \overline{ \cal F}^*(0)$, which can be written as
\begin{eqnarray}
S_{\rm eff} &=& 
\alpha_S (\gamma_{ST}-\gamma_{SL}) -\gamma  \\
&=& (\alpha_S-1) \gamma + \alpha_S S \nonumber
\end{eqnarray}
From the last equation it follows that the effective spreading coefficient
is always larger than the bare spreading coefficient $S$,
since $\gamma > 0$ and $\alpha_S > 1$.
The substrate area ratio can be expressed in terms
of the spreading coefficients and the liquid interfacial
tension as
\begin{equation}
\alpha_S= \frac{S_{\rm eff}+\gamma}{S+ \gamma}
\end{equation}

\subsection{Definition of roughness-induced wetting}
With the definitions of the last sections we are now able
to clearly define the subject and purpose
of the present work. As already stated in the
Introduction, we are concerned with the case where the flat substrate
is not wet, i.e., the free energy difference $\Delta {\cal F}^0(\ell)$ 
for the flat case, Eq. (6), is negative for some finite value of $\ell$.
The central question is: Under which conditions will the rough
substrate be wet, i.e., under which conditions does
\begin{equation}
\Delta \overline{\cal F}^*(\ell) >  0
\end{equation}
hold for all finite film thicknesses $\ell < \infty$? 
The answer of course imposes conditions 
both on the magnitude of substrate roughness (measured by $\alpha_S$)
and on the interactions between the coexisting phases, i.e., on
the molecular interaction $ w(\mbox{\boldmath $\rho$},z)$, 
which enters the calculation of the
wetting potential in (9).

In Section III.B we give two rather general necessary conditions
for roughness-induced wetting, which hold at very small
wetting-layer thickness and rather large layer thickness (as 
compared to the substrate roughness amplitude), respectively. 
For the intermediate
film thickness, we derive a sufficient condition in
Section III.C.

\subsection{Necessary conditions for roughness-induced wetting}
\subsubsection{Necessary condition for vanishing film thickness}

The  necessary condition for a roughness-induced wetting transition 
which corresponds to  (22) for vanishing film thickness 
($\ell \rightarrow 0$) follows from (19) and (20). It can be written
as $S_{\rm eff} > 0$, which together with the non-wetting
condition for the flat case ($S<0$) leads to the
inequalities\cite{Beaglehole2}
\begin{equation}
\gamma_{ST} - \gamma_{SL} < \gamma <  \alpha_S (\gamma_{ST}-\gamma_{SL})
\end{equation}
These inequalities can only
be satisfied if $\gamma_{ST} > \gamma_{SL}$
holds, since $\alpha_S \geq 1$ by definition.
From (5) one then also obtains that $\gamma >  -S$ has to hold.
Eq. (23) also imposes a lower bound on the substrate surface ratio
$\alpha_S$,
\begin{equation}
\alpha_S > \frac{1}{1+S/\gamma}
\end{equation}
Clearly, the conditions (23) or (24), 
although necessary, are not sufficient for wetting, 
 since the free energy $\Delta \overline{\cal F}^*(\ell)$
  can develop a minimum at finite separation
$\ell$. The conditions (23) and (24)
 simply correspond to a situation where the $\ell \rightarrow
\infty$ thick liquid layer has a lower interfacial energy than 
a film of vanishing thickness $\ell \rightarrow 0$.

\subsubsection{Harmonic approximation}

We now introduce the harmonic approximation, which we will adhere to
in the remainder of this paper.
Consider a corrugated solid surface, chosen
to have sinusoidal  undulations along one direction $\rho_1$
of the  two-dimensional reference plane $(\rho_1,\rho_2)$ with
amplitude $h_S$ and wave number  $q$,
\begin{equation}
\zeta_{S}(\mbox{\boldmath $\rho$}) \equiv
h_{S} \sin(q \rho_{1})
\end{equation}
The liquid profile is approximately (within linear response theory)
characterized by the same $q$-mode undulation with a different
amplitude $h_L$, vertically displaces by the film thickness $\ell$,
\begin{equation}
\zeta_{L}(\mbox{\boldmath $\rho$})\equiv
h_{L} \sin(q \rho_{1}) + \ell
\end{equation}
This geometry is depicted in Fig. 3.
To linear  order, there is also no phase shift between the two
surfaces. In the following, 
the amplitude $h_S$ is assumed to be positive, with no loss of generality.
The interfaces can not
penetrate each other, constituting the 
{\em non-crossing condition}, which can be written as
$\zeta_{L}(\mbox{\boldmath $\rho$}) 
\geq  \zeta_{S}(\mbox{\boldmath $\rho$})$, valid  at any point
$\mbox{\boldmath $\rho$}$.  This  leads to the
constraints
\begin{equation}
h_S - \ell \leq h_L \leq \ell + h_S
\end{equation}
Expanding the expressions for the interfacial
ratios $\alpha_S$ and $\alpha_L$, (14) and (15) and
keeping only terms up to quadratic order in the amplitudes
$h_L$ and $h_S$, leads to the following
expressions for the area ratios
\begin{equation}
\alpha_{S}=1+\frac{1}{4} h_S^2 q^2
+{\cal O}((h_Sq)^4)
\end{equation}
\begin{equation}
\alpha_{L}=1+\frac{1}{4} h_L^2 q^2
+{\cal O}((h_Lq)^4)
\end{equation}
which are expected to be good approximations for
weakly corrugated interfaces (as long as
$h_S q \ll 1$ and $h_L q \ll 1$).
The necessary condition 
for roughness-induced wetting at vanishing
film thickness (24)  becomes
\begin{equation}
\alpha_S \simeq 1+\frac{1}{4}h_S^2 q^2  > \frac{1}{1 + S/\gamma}
\end{equation}
We now identify two different physical regimes:
for $\gamma \stackrel{>}{\sim} -S$, defining the {\em interaction
dominated regime}, the necessary condition for wetting leads to
$h_S q \gg 1$ and the expansion in terms of $hq$ breaks down.
Here, the solid roughness has to be quite 
pronounced and  the behavior of the liquid interface turns out to be 
mostly dominated by the short-ranged part of the molecular interaction.
For $\gamma \gg -S$, defining the {\em tension dominated regime},
condition (30)  can be fulfilled even for 
small solid roughness ($h_S q \ll 1$); 
here, the liquid interface is dominated by its surface tension.
It is the tension-dominated  regime where the approximations
leading to (28) and (29) and other simplifications
made in the remainder this paper  are valid; this is also the regime 
of most practical interest, since values for $h_S q$ characterizing
rough surfaces in experiments are typically quite small\cite{estimate}.

\subsubsection{Necessary condition for thick films}

An additional  necessary  condition for wetting is 
$\overline{P}^*(\ell)>0$ valid for average film 
thicknesses approximately larger
than the corrugation amplitude, $ \ell \stackrel{>}{\sim} h_S$.
This condition  can be obtained in the following way:
suppose we have a flat liquid interface, i.e., $\alpha_L=1$.
Then the first term in (18) vanishes, and in order
for the minimized free energy difference 
$\Delta \overline{\cal F}^*(\ell)$ to be
positive we have to require  $\overline{P}^*(\ell)>0$.
Clearly, a flat liquid interface is only possible 
for a liquid layer thickness which is larger than the
amplitude of the solid roughness, otherwise the two interfaces
sterically interact. Figure 4 schematically depicts the
limiting case $\ell \approx h_S$ with the flat liquid
interface just touching the solid substrate at the largest
height fluctuation characterized by the amplitude $h_S$.
For sinusoidal interfaces described by (25) and (26) and
using  the non-crossing constraint (27),
one obtains the inequality
\begin{equation}
\overline{P}^*(\ell)>0 \;\;\;\; \mbox{for} \;\; h_S < \ell < \infty
\end{equation}
For smaller distances,
the interaction $\overline{P}^*(\ell)$ can actually become negative with
$\Delta \overline{\cal F}^*(\ell)$ still being strictly positive, 
because then the free energy expression (18) always has a positive
energy contribution from the interfacial tension
of the liquid interface.

This result has consequences for the important class of wetting
potentials with a single minimum at finite but rather large
wetting film thickness, which describe continuous wetting transitions
as the minimum moves outwards to infinity:  
if the minimum occurs at distances larger 
than  the roughness amplitude, it follows from (31) that  the substrate
roughness will not induce the wetting of the substrate.

From the above considerations  we see that the interaction
$\overline{P}^*(\ell)$ has to have rather complex behavior; 
at zero $\ell$ or vanishing
liquid film it is negative, since $\overline{P}^*(0) \approx  \alpha_S S$
and we start with the assumption of a 
non-wetting behavior (i.e., $S<0$)  for the flat solid surface. 
Only considering the $\ell=0$ situation we see that there
is a threshold value  of the solid roughness
in order to make the vanishing film limit energetically
unfavorable compared to the infinite film limit,
see (24). For a film
thickness larger than the amplitude of the solid roughness,
the interaction $\overline{P}^*(\ell)$  has to be repulsive in order to make 
roughness-induced wetting possible, see (31).
The question that arises naturally is whether such a 
behavior is actually possible and what the conditions on 
the molecular interaction $w({\bf r})$  are.

\subsection{Sufficient condition for roughness-induced wetting}

In this section, we want to show for general film thicknesses
under which conditions roughness-induced wetting, as defined
by Eq. (22), occurs.
In order to do so, we need to minimize the free energy
with respect to the fluid interface
profile $\zeta_L(\mbox{\boldmath $\rho$})$
 for each average film thickness $\ell$ and for
a given molecular interaction $w(\mbox{\boldmath $\rho$},z)$
and interfacial tensions, according to (16) [with 
$\zeta_L(\mbox{\boldmath $\rho$})$ entering the expressions (8) and (9)].
We then have to check whether $\Delta \overline{\cal F}^*(\ell)>0$
holds for each $\ell$. If it turns out
that $\Delta \overline{\cal F}^*(\ell)<0$
for a given $\ell$, we know that the system will prefer to have
a stable film of this thickness, and wetting will not occur.

For the present purpose, we turn this procedure somewhat around: we will 
try to determine  the interaction for which roughness-induced wetting,
for given values of the interfacial tensions and the substrate roughness, 
does occur. Furthermore, in
many situations the molecular interaction $w(\mbox{\boldmath $\rho$},z)$
is not easily available, and usually the wetting potential
for two flat interfaces, $P^0(\ell)$, as defined by (2),
is readily measured and calculated. We will therefore use a novel
relation between the planar interaction $P^0(\ell)$ and the wetting
potential between two sinusoidally modulated interfaces, 
$P(\rho_1, \ell,  h_L)$,
which is derived in the appendix.
Using this relation, we express the condition for roughness-induced wetting
in terms of a lower bound on the planar interaction $P^0(\ell)$.
The analysis will be presented in the next two subsections.
These parts are somewhat technical, and the unmotivated
reader can easily skip these paragraphs and move on to the
results in Section III.C.3.

\subsubsection{Construction of lower bound  on $P^0(\ell)$}

To proceed, consider first the range $\ell < h_S$.
In the tension-dominated regime,
defined  by $h_S q \ll 1$, it follows that $\ell q \ll 1$ also holds.
In this limit, the  wetting potential, defined by (9), can for
the special case of two sinusoidal interfaces (see Section III.B.2)
be expressed in terms of the planar interaction
$P^0(\ell)$. Neglecting
curvature-like terms which turn out to scale like $h_S h_L  q^4$,
this  relation is given by (see Appendix)
\begin{equation}
P(\rho_1,\ell,h_L) \simeq 
(1+h_S h_L q^2 \cos^2[q \rho_1])^{1/2} P^0 \left(
\frac{\ell+(h_S-h_L)\sin[q \rho_1]}{
(1+h_S h_L q^2 \cos^2[q \rho_1])^{1/2}} \right)
\end{equation}
Clearly, the above form has the following desired property:
for $\ell =0$ one has $h_L = h_S$ and one thus 
obtains for the spatially averaged potential
$\overline{P}(\ell=0,h_L) \simeq  \alpha_S P^0(\ell=0)$,
as anticipated on intuitive grounds  in the previous section.
For either $h_S=0$ or $h_L=0$ the formula (32) simplifies to
$P(\rho_1,\ell,h_L)=P^0 (\ell+(h_S-h_L)\sin[q \rho_1])$, which is 
exact.

Still, the formula (32) is rather complicated and calculating
the spatially averaged potential 
$\overline{P}(\ell,h_L) \equiv 
\langle P(\rho_1,\ell,h_L) \rangle_{\rho_1}$
is by no means trivial.
The following observation is crucial: for a given
value of $\ell$, the argument of $P^0$ in (32) is strictly
smaller than $2 \ell$ (since $2 h_S \ell q^2 < 1$).
We can therefore choose  a linear trial function of the form
\begin{equation}
P^0_T(t)=S+ c(\ell) t
\end{equation}
For a given value of $\ell$,
we require the trial $P^0_T(t)$ to be a lower bound of the
flat-interface potential $P^0(t)$, i.e., $P^0_T(t)< P^0(t)$, 
in the finite range $0 \leq t \leq 2 \ell$, which puts
certain bounds on the slope $c(\ell)$.
It trivially  follows that $P(\rho_1,\ell,h_L)$ as given 
by (32) and evaluated with 
$P^0_T$ is always lower than $P(\rho_1,\ell,h_L)$ 
evaluated with $P^0$ instead.

The strategy will be as follows:
If we can show that for a film with a given average thickness $\ell$ and
using the linear trial function $P^0_T$ the resulting free energy 
difference $\Delta \overline{\cal F}^*(\ell)$ is positive, 
we know that the wetting film for this particular
mean thickness is unstable.  
This follows since using $P^0$ instead of $P^0_T$ we will necessarily
increase $\overline{P}^*(\ell)$ and 
thus also increase $\Delta \overline{\cal F}^*(\ell)$.
If we succeed in showing the same  
for all values of $\ell$ (including $\ell=0$),
it follows that $P^0(\ell)$ (which is by construction
strictly larger than all the linear
trial functions) is an interaction which shows roughness-induced wetting.
In the subsequent calculations we actually revert this procedure and
start with the linear trial functions, enforcing instability of the
wetting film for each value of $\ell$ separately [leading to
restrictions on the $\ell$-dependent slopes $c(\ell)$], and construct the
function $P^0(\ell)$ from a superposition of all piece-wise linear 
functions.  Choosing a linear trial function of the form (33)
does not restrict the generality of our results,
since one can express any function as the supremum of
a set of suitable piecewise linear functions.

Using the trial function (33) the averaging 
over the coordinate $\rho_1$ can be easily done
and,with (18) and (32), the  resultant  trial free energy is given by
\begin{equation}
\Delta \overline{\cal F}_T(\ell,h_L)=
\frac{1}{4} \gamma h_L^2 q^2 + S(1+\frac{1}{4} h_S h_L q^2 ) + c(\ell)  \ell
+{\cal O}(q^4)
\end{equation}
Minimizing this free energy expression with respect to 
$h_L$, one obtains the following amplitude
\begin{equation}
h_L^*=- \frac{S}{2 \gamma} h_S
\end{equation}
which characterizes the minimizing liquid interface profile 
$\zeta_L^*$.
The liquid interface corrugation is in phase with the substrate
configuration, since $S<0$.
Since the two interfaces have to satisfy the non-crossing 
constraint (27),
the minimizing  amplitude $h_L^*$  can be realized only for
a thickness larger than some characteristic length $\ell \geq \ell'$.
For $\ell=\ell'$, the liquid interface has an amplitude
given by (35) and touches the substrate surface; this situation
is depicted in Fig. 5. 
This defines the length $\ell'$, which is, combining (27) and (35),  
given by
\begin{equation}
\ell'= h_S(1+\frac{S}{2 \gamma})
\end{equation}
For smaller film thicknesses $\ell$, the amplitude
$h_L$ necessarily deviates from the value $h_L^*$ and finally
equals $h_S$ for $\ell=0$. 
Note that the length $\ell'$  is always bounded
by $h_S/2 < \ell' < h_S$, as follows from the fact that
$\gamma > -S$, see (23).

\subsubsection{Calculation for $S_{\rm eff}=0$}

In this paragraph we 
assume that $S_{\rm eff}=0$, that means the substrate
roughness just suffices to make the non-wetting  state ($\ell=0$)
energetically unfavorable compared to the completely
wet state ($\ell= \infty$) at coexistence ($\mu=0$).
This is obviously not necessary but renders the resultant
expressions in a simpler form and 
is sufficient for $\mu=0$.
The extension to general $S_{\rm eff}$ will be done
in Section III.C.4.
From (30) the excess substrate surface area $q^2 h_S^2/4$
is given by
\[
\frac{q^2 h_S^2}{4} = -\frac{S}{S+\gamma} 
\]
For $\ell <  \ell'$, the amplitude which minimizes
the free energy and is in accord with the non-crossing  constraint
is given by $h_L=h_S-\ell$, that is, the interfaces just touch.
Inserting this amplitude into the trial free energy expression (34),
one obtains the minimal value (denoted by an asterisk)
\begin{equation}
\Delta \overline{\cal F}^*_T(\ell)=
c(\ell)  \ell 
-\frac{\ell^2}{h_S^2} \left(
\frac{\gamma S}{S+ \gamma} \right)
+ \frac{S \ell}{h_S} \left( 
\frac{S+2 \gamma }{S+\gamma} \right)
\;\;\;\;\mbox{for}\;\;\;\; \ell<\ell'
\end{equation}
Thus, the sufficient  condition for 
the free energy difference to be positive is 
\begin{equation}
c(\ell) \geq  - \frac{S }{h_S} 
\left( \frac{S+ 2 \gamma }{S+\gamma} \right)
+\frac{\ell}{h_S^2} \left(
\frac{\gamma S}{S+ \gamma} \right)
\;\;\;\;\mbox{for}\;\;\;\; \ell<\ell'
\end{equation}
with the special values
\begin{equation}
c(0) \geq 
  - \frac{S }{h_S} \left( \frac{S+ 2 \gamma }{S+\gamma} \right)
\end{equation}
and 
\begin{equation}
c(\ell')  \geq 
  - \frac{S }{2 h_S} \left( \frac{S+ 2 \gamma }{S+\gamma} \right)
\end{equation}
For $\ell >   \ell'$, one inserts the expression (35)  found
for  $h_L^*$ into the free energy (34) and obtains
\begin{equation}
\Delta \overline{\cal F}^*_T(\ell ) =
c(\ell)  \ell + S - 
\frac{h_S^2 S^2 q^2}{16 \gamma}
\;\;\;\;\mbox{for}\;\;\;\; \ell>\ell'
\end{equation}
Using  that $S_{\rm eff}=0$,
it follows that 
\begin{equation}
\Delta \overline{\cal F}^*_T(\ell ) =
c(\ell)  \ell + S + 
\frac{S^3}{4 \gamma(S+\gamma)}
\;\;\;\;\mbox{for}\;\;\;\; \ell>\ell'
\end{equation}
In this case, requiring 
the free energy to be positive leads to the condition 
\begin{equation}
c(\ell ) \geq  - \frac{S }{4 \gamma \ell } 
 \frac{(S+ 2 \gamma)^2 }{S+\gamma} 
\;\;\;\;\mbox{for}\;\;\;\; \ell>\ell'
\end{equation}
with the limiting value (40) for $\ell=\ell'$.

Thus far, we have calculated a lower bound of the
planar potential $P^0(t)$ for a given value of the
average film thickness $\ell$, which
consists of a linear function of finite extent (between
$0$ and $2 \ell$) and $\ell$-dependent slope.
In order to construct a lower bound on $P^0(t)$ which is valid
for {\em any} film thickness, we have to
calculate the upper  envelope of these linear functions. 
This function we denote by $P^0_{\rm low}(\ell)$. The
sufficient condition for wetting, corresponding to
the definition (22), can then be written as 
\begin{equation}
 P^0(\ell) >  P^0_{\rm low}(\ell)
\end{equation}
with $P^0_{\rm low}(\ell)$ being defined as
\begin{equation}
P^0_{\rm low}(\ell) \equiv \max_{t \geq \ell/2} \{S + c(t) \ell\}
\end{equation}
The value of $c(t)$ is given by (38) and (43) for $t<\ell'$ and $t>\ell'$,
respectively.
Since $c(t_1)>c(t_2)$ holds for any $t_1< t_2$, it is easy to see that
the function $P^0_{\rm low}(\ell)$ can be expressed in closed form as
\begin{equation}
P^0_{\rm low}(\ell) =
S+c(\ell/2) \ell
\end{equation}
Using the expressions (38) and (43), the 
function $P^0_{\rm low}(\ell)$ is explicitly given by
\begin{equation}
P^0_{\rm low}(\ell)  = \left\{ \begin{array}{lll}
        S \left( 1-\frac{\ell}{h_S} \frac{S+2\gamma}{S+\gamma}
       + \frac{\ell^2}{h_S^2} \frac{\gamma/2}{S+\gamma} \right)

         & \mbox{for }     & \ell \leq  2\ell'  \\
        S \left( 1-\frac{(S+2\gamma)^2}
        {2\gamma(S+\gamma)} \right)   & \mbox{for }  & 2\ell' \leq \ell \leq  \ell_{\rm max} \\
        0  & \mbox{for }  & \ell > \ell_{\rm max} \\
                \end{array} \right.
\end{equation}

\subsubsection{Results for $S_{\rm eff}=0$}

Figure 6 shows  the rescaled lower bound 
$-P^0_{\rm low}/S$ as a function of the rescaled
variable $\ell/h_S$ for $-S/\gamma=1/2$. 
This  function starts out at $-1$ for $\ell/h_S=0$ by definition of the
spreading coefficient $S$. The slope at the origin is finite,
and the function increases until it reaches its maximum at $\ell=2\ell'$.
We obtain a plateau in the interval $2\ell'<\ell<\ell_{\rm max}$; for
$\ell > \ell_{\rm max}$ the lower bound is given by $-P^0_{\rm low}(\ell)/S=0$.
In our estimate for the value of $\ell_{\rm max}$ we set
$\ell_{\rm max}=\ell^* + 2 h_S$, where $\ell^*$ is defined by
$P_{\rm low}^0(\ell^*)=0$ (see Fig. 6).
This is based on the observation that for $\ell > \ell_{\rm max}$ 
the interaction as given by (32) is strictly positive as 
long as $|h_L|<h_S$, since in this case the arguments of 
$P^0$ in (32) are always larger than $l^*$;
consequently, the free energy difference (18) will be always  positive.
If, on the other hand, $|h_L|>h_S$ holds, the free energy
difference (18) will  also be positive since  then 
$\alpha_L(\ell,h_L)>\alpha_L(0)$ and 
$\overline{P}^*(\ell)>\overline{P}^*(0)$ holds 
for any  $\ell> 0$.

The value for $\ell^*$ as obtained from (47) is given by
\begin{equation}
\frac{\ell^*}{h_S}=
\frac{S+2 \gamma}{\gamma} -
\sqrt{\left( \frac{S+2 \gamma}{\gamma} \right)^2
-\frac{2(S+ \gamma)}{\gamma}} 
\end{equation}
and reduces to the value $\ell^*/h_S \approx 2- \sqrt{2} \approx 0.59$
in the tension-dominated regime,
$\gamma \gg -S$, and approaches zero 
in the interaction dominated regime,
 $\gamma \stackrel{>}{\sim} -S$.
The largest possible value for $l_{\rm max}/h_S $ is thus
 $l_{\rm max}/h_S  = 4- \sqrt{2} \approx 2.59$.
In the tension-dominated regime,
 $\gamma \gg -S$,
the assumption we started with,
$\ell q \ll 1$,
thus holds for the whole range of
thicknesses considered because the substrate roughness
necessary to achieve roughness-induced wetting is small,
$h_S q \ll 1$. 
In the same limit, the general expression
(47)  takes the values
\begin{equation}
P^0_{\rm low}(\ell)  \approx  \left\{ \begin{array}{lll}
        S \left( 1-2 \frac{\ell}{h_S} 
       + \frac{1}{2} \frac{\ell^2}{h_S^2}  \right)

         & \mbox{for }     & \ell/h_S  \leq  2\\
        -S
        & \mbox{for }  & 2\leq  \ell/h_S  \leq  4- \sqrt{2} \\
        0  & \mbox{for }  & \ell/h_S  > 4- \sqrt{2} \\
                \end{array} \right.
\end{equation}

In the interaction dominated regime, for 
 $\gamma \stackrel{>}{\sim} -S$, the maximum of 
$P^0_{\rm low}(\ell)$ for $ 2\ell' \leq  \ell \leq \ell_{\rm max}$ actually diverges
as $\gamma \rightarrow -S$, with $2\ell'/h_S \rightarrow 1$ and
$\ell_{\rm max}/h_S  \rightarrow 2$. Note that in this regime, also the
substrate roughness necessary to induce wetting goes to infinity, see (24).
It  follows that it is the tension-dominated regime where roughness-induced
is most likely to be observed experimentally.

The function $P^0_{\rm low}(\ell)$ as given by (47) 
is plotted in Fig. 7
for $-S/\gamma=0$, $-S/\gamma =1/2$ 
and $-S/\gamma =2/3$, from bottom to top.
As this ratio  becomes larger, as one moves from the tension-dominated
regime into the interaction-dominated regime, the maximum
of $-P^0_{\rm low}(\ell)/S$ increases, until it finally reaches infinity 
for $-\gamma/S=1$. 

The interpretation of the results in Fig. 7 is the following: 
For a given ratio of the spreading coefficient $S$ and the liquid
interfacial tension $\gamma$, which are  determined by the interfacial
tensions of the problem alone and conspire to
give $S_{\rm eff}=0$, one rescales the amplitude of the wetting potential 
$P^0(\ell)$ (which can be measured or calculated theoretically)
by the spreading coefficient and the distance $\ell$ by the 
roughness amplitude $h_S$. If the rescaled potential 
$-P^0(\ell/h_S)/S$ is larger than $-P_{\rm low}^0(\ell/h_S)/S$ for
all arguments, i.e. if the sufficient condition (44) is fulfilled, 
a complete wetting situation will result. 

An interesting feature of our results comes from the fact that we 
obtain universal lower bounds as a function of the film thickness
$\ell$ rescaled by the roughness amplitude $h_S$.
This suggests that there exists for a given substrate excess surface area
$\sim h_S^2 q^2 /4$ an optimal value of the roughness amplitude
$h_S$ for roughness-induced wetting to occur.
Since the maximum of the lower bound $-P_{\rm low}^0(\ell/h_S)/S$ is located 
at $\ell/h_S \approx 1$, this optimal value of $h_S$ happens to 
coincide with the approximate 
location of the maximum of $P^0(\ell)$, the wetting potential.
Assume that the sufficient condition (45) is in fact satisfied for 
this optimal value of $h_S$. For roughness amplitudes much smaller than this
optimal value (the function $P^0$ will be pushed to the 
right in the scaling plot 
Fig. 7) the sufficient condition cannot be satisfied for very small
film thicknesses due to the finite slope of the wetting potential, 
suggesting an infinitesimally thin stable film (no
wetting). For roughness amplitudes much larger than the optimal value
(the function $P^0$ will be squeezed to the left in Fig. 7)
the sufficient condition is not satisfied for average 
film thicknesses larger than some value at the order of the roughness
amplitude, suggesting the formation of a very thin liquid film
occupying preferentially the valleys of the substrate surface
fluctuations (partial prewetting).

\subsubsection{Results for general $S_{\rm eff}$}

Here we treat the general case with $S_{\rm eff}>0$,
now including the cases  where the substrate roughness $\alpha_S$
is larger than the necessary value obtained by setting 
$S_{\rm eff}=0$ in (21).
In addition, we will require the free energy difference (18) 
to be larger than a constant denoted by ${\cal F}_0$,
as will turn out to be important if one looks at
stable wetting layers in the presence of a chemical 
potential (see Section III.D).
Using a trial function of the form (33) the following
bounds for the slope $c(\ell)$ follow

\begin{equation}
c(\ell) \geq   \frac{1 }{h_S}
\frac{(S+ 2 \gamma)(S_{\rm eff}-S)}{S+\gamma} 
-\frac{\ell}{h_S^2}
\frac{\gamma (S_{\rm eff}-S)}{S+ \gamma} 
+\frac{{\cal F}_0-S_{\rm eff}}{\ell}
\;\;\;\; \mbox{for} \;\;\;\; \ell < \ell'
\end{equation}
\begin{equation}
c(\ell ) \geq   \frac{{\cal F}_0}{\ell}
+ \frac{S^2 S_{\rm eff}-S(S+ 2 \gamma)^2 }{4\gamma \ell(S+\gamma)}
\;\;\;\; \mbox{for} \;\;\;\; \ell > \ell'
\end{equation}
Since $c(\ell )$ for $\ell < \ell'$
 now is non-monotonic and has a maximum
at $\overline{\ell}$ given by
\begin{equation}
\overline{\ell}=h_S \sqrt{\frac{(S+\gamma)(S_{\rm eff}-{\cal F}_0)}
{\gamma (S_{\rm eff}-S)}}
\end{equation}
the global lower bound $P^0_{\rm low}(\ell)$ can be calculated according 
to (46) only in the restricted range of $2 \overline{\ell}< \ell< 2\ell'$.
For $\ell < 2 \overline{\ell}$, the function $P^0_{\rm low}(\ell)$ as defined
by (45) has a constant slope of  $c(\overline{\ell})$
given by (50)  since $\overline{\ell}< \ell'$
strictly holds.
The global lower bound is thus given by

\begin{equation}
P^0_{\rm low}(\ell)  = \left\{ \begin{array}{lll}
        S+\frac{\ell}{h_S(S+\gamma)} \left(
        (S_{\rm eff}-S)(S+ 2 \gamma)-
        2\sqrt{\gamma (S_{\rm eff}-{\cal F}_0)(S_{\rm eff}-S)
        (S+\gamma)}\right) 
         & \mbox{for }     & \ell \leq  2\overline{\ell}  \\
        S +2({\cal F}_0-S_{\rm eff})
       +\frac{\ell}{h_S} \frac{(S+2\gamma)(S_{\rm eff}-S)}{S+\gamma}
       - \frac{\ell^2}{h_S^2} \frac{\gamma(S_{\rm eff}-S)}
       {2(S+\gamma)} 
         & \mbox{for }     &  2\overline{\ell}  \leq \ell \leq  2\ell'  \\
        S +2 {\cal F}_0+\frac{S_{\rm eff}S^2-S(S+2\gamma)^2}
        {2\gamma(S+\gamma)}  

 & \mbox{for }  & 2\ell' \leq \ell \leq  \ell_{\rm max} \\
        {\cal F}_0  & \mbox{for }  & \ell > \ell_{\rm max} \\
                \end{array} \right.
\end{equation}
In Fig. 8 we plot $P^0_{\rm low}(\ell)$ for the fixed value
$-S/\gamma=1/2$ and the values $S_{\rm eff}=0$,
$S_{\rm eff}/S=-1$, and $S_{\rm eff}/S=-2$ (from bottom to top in the 
right portion in the graphs).
The  interesting feature is that the plateau of
$P^0_{\rm low}(\ell)$ for $\ell > 2\ell'$ increases
with increasing $S_{\rm eff}$. This indicates
that for a given wetting potential $P^0(\ell)$
and a given roughness amplitude $h_S$ 
the roughness-induced wetting transition
might disappear for wave numbers  much 
larger than the threshold value determined by (30).

\subsection{Stable films in the presence of a chemical potential}

So far, we have calculated a lower bound for the interaction,
denoted by $P^0_{\rm low}(\ell)$, so that the free energy  
difference $\Delta \overline{\cal F}^*(\ell)$
is positive, Eq. (37), or larger than a positive constant
${\cal F}_0$, Eq. (43), for all values of $\ell$.
If $P^0(\ell) \geq P^0_{\rm low}(\ell)$ holds and the 
chemical potential vanishes, i.e., at coexistence, 
the substrate will be covered with an infinitely thick liquid 
layer.
For non-zero chemical potential, however, the substrate will
always be covered with a film of {\em finite} thickness.
For clarity sake, let us assume the potential $P^0(\ell)$ to 
have a single maximum at finite separation, decay for
large separations like $P^0(\ell) \sim a \ell^{-\sigma}$
(where $\sigma=m-4$ if the molecular interaction is given
by (A7))
and be negative for zero separation, as corresponding to the
non-wetting situation for a flat substrate. One notes 
that $\sigma=2$ for non-retarded van der Waals interactions.

For finite chemical potential, the free energy then has a minimum 
at a finite separation $\ell_{min}$.
Assuming that $\ell_{min} \gg q^{-1}$, one can use 
Eq. (A21) and finds the leading terms of
the free energy according to (18)
\begin{equation}
\Delta \overline{\cal F}(\ell,h_L)=\frac{\gamma}{4} h_L^2 q^2 +P^0(\ell)
+\frac{1}{4}(h_S^2 + h_L^2)P^{II}(\ell) +\mu \ell
\end{equation}
where the chemical potential term  has been added.
For a potential of the asymptotic form $\sim a \ell^{-\sigma}$
this free energy is minimized by $h_L=0$ and the 
film thickness which is stable with respect to variations in $\ell$
is asymptotically given by
\begin{equation}
\ell_{min} \sim (a \sigma/ \mu)^{1/(1+\sigma)}
\end{equation}
Choosing the constant ${\cal F}_0$ in (53) to be
${\cal F}_0=\Delta \overline{\cal F}^*(\ell_{min})=
\Delta \overline{\cal F}(\ell_{min},h_L=0)$ and 
demanding $P^0(\ell)>P^0_{\rm low}(\ell)$ as given by (53),
the minimum at $\ell_{min}$ is indeed the global minimum
and we have a stable wetting film of finite thickness.
We will now do a local stability 
analysis for this free energy minimum.
The linearized Euler equation, determining the
stable liquid interface profile,  can be
obtained from the free energy expression (8)
by requiring local force equilibrium, $\partial
{\cal F}(\mbox{\boldmath $\rho$},[\zeta_L])/
\partial \zeta_L(\mbox{\boldmath $\rho$})=0$,
and is given by
\begin{equation}
\gamma \triangle \zeta_L(\mbox{\boldmath $\rho$})
+\Pi^0(\ell) -\int \mbox{d}^2\mbox{\boldmath $\rho'$} \left\{
 \zeta_L(\mbox{\boldmath $\rho$})-
\zeta_S(\mbox{\boldmath $\rho+\rho'$})-\ell \right\}
w(\mbox{\boldmath $\rho'$},\ell) =\mu 
\end{equation}
where $\Pi^0(\ell) \equiv -\mbox{d}  P^0(\ell)/\mbox{d} \ell$ is the
disjoining pressure for flat interfaces.
Assuming the wetting layer to have a thickness
corresponding to the  global minimum as given by (55),
which is equivalent to setting $\Pi^0(\ell_{min})=\mu$,
and approximating the interfacial profiles again by
sinusoidal waves, Eqs. (25) and (26),
the locally stable liquid interface amplitude turns 
out to be\cite{Andelman} 
\begin{equation}
h_L= \frac{h_S \int \mbox{d}^2\mbox{\boldmath $\rho$} \;\;
\cos(q \rho_1) w(\mbox{\boldmath $\rho$},\ell)}{
\gamma q^2 + \int \mbox{d}^2\mbox{\boldmath $\rho$} \;\;
 w(\mbox{\boldmath $\rho$} ,\ell)}
\end{equation}
Using formulas (A9) and (A25), this amplitude
scale like
$h_L \sim h_S \exp(-q \ell) \ell^{1/2-m/2} q^{m/2-7/2}$
 in the limit $q\ell \gg1$;
for van der Waals forces (m=6) one obtains
$h_L \sim h_S \exp(-q \ell) \ell^{-5/2}  q^{-1/2}$.
In contrast to the global 
stability analysis leading to (55),
where we  averaged  over the
spatial coordinate \(\mbox{\boldmath $\rho$}\),
the local equilibrium analysis now actually gives
a non-vanishing amplitude $h_L$.
Inserting this amplitude back  into the free energy 
expression (54), the stable film thickness
is increased. However, the correction turns out
to be less singular for the case of van der Waals forces and thus
does not affect the asymptotic behavior of $l_{min}$
as given by (55).

\section{Discussion}
Some necessary conditions for roughness-induced wetting have been
obtained on very general grounds. 
First, the tension of the liquid interface has to be larger
than the negative spreading coefficient, or, equivalently,
the tension of the substrate-vapor interface has to be larger
than the tension of the substrate-liquid interface.
Second, the roughness of the substrate, measured by the
excess area as compared to the flat substrate, has to 
exceed a certain threshold, see (24).
Using an approach which is equivalent to a linear response analysis,
we obtain a formula relating the molecular interaction
between two rough interfaces to the
interaction of two flat interfaces; the latter
is the so-called wetting potential.
Using this formula,
a lower bound for wetting potentials in order to obtain 
this new type of wetting transition  is derived.
This lower bound constitutes a sufficient condition.
More specifically, this lower bound has the following
properties:  i) for vanishing liquid film, or for
zero separation between the two interfaces bounding
the liquid layer, the interaction (which for this limit
is the so-called spreading
coefficient) is negative, corresponding  to
a non-wetting situation
in the case of flat interfaces;
ii) the lower bound has a pronounced 
maximum with a height of
at least the negative spreading  coefficient at
a separation of about  the roughness amplitude.
This height depends on the ratio between the liquid-interface
tension $\gamma$  and the spreading coefficient $S$.

These results concerning the possibility
of a roughness-induced wetting transition
agree with the very recent experimental
observation of 
premelting induced by substrate roughness
at the glass-ice interface\cite{Beaglehole2},
where glass surfaces showing a roughness with 
typical amplitudes of $\sim 1 nm$ induced a complete interfacial 
premelting. Independently, the calculation of the dispersion interaction
between half spaces of glass and ice separated by a 
liquid water layer show a pronounced maximum at about the
same distance $\sim 1nm$\cite{Wilen},
consistent with  our prediction for the 
lower bound of the wetting potential.
We find the asymptotic behavior
of the film thickness as a function of the chemical
potential to be  characterized by the standard van
der Waals exponent,
as shown in Sec. III.D. This is in disagreement with
the experimentally measured exponent $\simeq 2$\cite{Beaglehole2},
which is to be contrasted with the exponent 
$1/3$ as expected for non-retarded van der Waals
interactions in the absence of any other interactions. 
Finally, we  note that the interfacial energies for the
case of interfacial premelting are influenced  by grain-boundary
energies in the disordered polycrystalline layers adjacent
to the microscopically irregular substrate wall.
In fact, it was argued \cite{Dash3} that this effect 
could independently lead to a roughness-induced premelting transition.
Such a mechanism could be described within our framework 
by assuming an interfacial energy $\gamma$ which depends
explicitly on the roughness magnitude of the liquid interface.
A further interesting effect might appear for the case
of surface triple-point premelting: here a roughness-induced
complete surface melting could be triggered by a roughening 
transition of the solid surface\cite{Dietrich4}.
Another interesting consequence of our results is that 
a discontinuous wetting transition  might be converted
into a continuous transition by a change of the effective
wetting potential at small distances. 

Our  calculations are most valid for 
small substrate roughness, as defined by $h_S q < 1$;
this also seems to be the experimentally most relevant
limit, since rough surfaces produced by etching usually
show modest corrugation amplitudes\cite{estimate}.
The  closed-form expressions for the interaction of two corrugated
interfaces obtained in the Appendix
are to the best of our knowledge novel
and applicable to a wide range of phenomena,
including dewetting  phenomena (where the whole argument has to be
inverted, leading to roughness-induced dewetting).
We also derive an  expression 
for the curvature contribution to the free  energy, which might play
a role in the 
adsorption of
membranes or vesicles on rough substrates. 

\acknowledgments
This project initiated when one of us (DA) visited
the laboratory of D. Beaglehole.
He is grateful to D. Beaglehole for introducing him to the
subject of roughness-induced wetting, for sharing with him his
unpublished experimental results, and for his hospitality.
DA acknowledges partial support from the German Israel
Foundation (GIF) under grant No. I-0197 and
the US-Israel Binational Science Foundation (BSF)
under grant No. 94-00291.
RN acknowledges support from the Minerva Foundation,
receipt of a NATO stipend administered by the DAAD, and
partial support by the National Science Foundation under
Grant No. DMR-9220733.
We would like to thank G. Dash, S. Dietrich, M. Elbaum, F. Joanny, 
S. Safran, 
M. Schick, U. Steiner, J. Wettlaufer, and L. Wilen
for helpful discussions.

\appendix
\section{Derivation of the effective potential}
The total interaction per unit (projected) area between the liquid interface
at lateral position \(\mbox{\boldmath $\rho$}\) 
located at 
\(\zeta_{L}(\mbox{\boldmath $\rho$})\), and
a corrugated solid surface, parameterized by 
\( \zeta_{S}(\mbox{\boldmath $\rho$}) \) (see Fig. 3),
can be written as
\begin{equation}
P(\mbox{\boldmath $\rho$},[\zeta_{L}])=
\int_{\zeta_{L}(\mbox{\boldmath $\rho$})}^{\infty} \mbox{d}z
\int \mbox{d}^{2}\mbox{\boldmath $\rho'$}
\int_{-\infty}^{\zeta_{S}(\mbox{\boldmath $\rho +\rho'$})} \mbox{d}z' \;\;\;
w(\mbox{\boldmath $\rho'$},z-z')
\end{equation}
This expression is hard to deal with due to the non-local dependence
on the shape of the solid surface, which enters in the integration boundary.
Progress can be made by formally expanding \( \zeta_{S}(\mbox{\boldmath 
$\rho +\rho'$})\) around 
\(\zeta_{L}(\mbox{\boldmath $\rho$}) -\ell\).
The average separation between the two interfaces, $\ell$, is given by
\( \ell \equiv \langle \zeta_{L}(\mbox{\boldmath $\rho$})
-\zeta_{S}(\mbox{\boldmath $\rho$})\rangle\), where the
brackets denote a spatial average over \( \mbox{\boldmath $\rho$}\). 

Keeping terms up to fourth order, one obtains
\begin{eqnarray}
P(\mbox{\boldmath $\rho$},[\zeta_{L}]) &=&
P^{0}(\ell) + 
\int_{\ell}^{\infty} \mbox{d}z \int \mbox{d}^{2}\mbox{\boldmath $\rho'$}
\left( \zeta_{S}(\mbox{\boldmath $\rho+\rho'$})-\zeta_{L}
(\mbox{\boldmath $\rho$})+\ell
\right) w(\mbox{\boldmath $\rho'$},z) +  \nonumber \\
&& \frac{1}{2}
\int \mbox{d}^{2}\mbox{\boldmath $\rho'$}
\left( \zeta_{S}(\mbox{\boldmath $\rho +\rho'$})-
\zeta_{L}(\mbox{\boldmath $\rho$})+\ell
\right)^2w(\mbox{\boldmath $\rho'$},\ell) + \nonumber \\
&& \frac{1}{6} \left.
\int \mbox{d}^{2}\mbox{\boldmath $\rho'$}
\left( \zeta_{S}(\mbox{\boldmath $\rho+\rho'$})-\zeta_{L}
(\mbox{\boldmath $\rho$})+\ell
\right)^3 \frac{\partial}{\partial z}w(\mbox{\boldmath $\rho'$},z) 
\right|_{z=\ell} + \nonumber \\
&& \frac{1}{24} \left.
\int \mbox{d}^{2}\mbox{\boldmath $\rho'$}
\left( \zeta_{S}(\mbox{\boldmath $\rho+\rho'$})-
\zeta_{L}(\mbox{\boldmath $\rho$})+\ell
\right)^4 \frac{\partial^2}{\partial z^2}w(\mbox{\boldmath $\rho'$},z)
\right|_{z=\ell}
\end{eqnarray}
where the expression for the interaction of  two planar
surfaces introduced in (2),
\begin{equation}
P^{0}(\ell) \equiv
\int_{\ell}^{\infty} \mbox{d}z
\int \mbox{d}^{2}\mbox{\boldmath $\rho$}
\int_{-\infty}^{0} \mbox{d}z' \;\;\;	
w(\mbox{\boldmath $\rho$},z-z')
\end{equation}
has been used.
At this point it is useful to specify the interfacial profiles;
we choose  one-dimensional sinusoidal profiles 
for the liquid and the solid surfaces, as depicted in Fig. 3,
as is sufficient and appropriate for a linear analysis,
\begin{equation}
\zeta_{S}(\mbox{\boldmath $\rho$}) \equiv
\zeta_{S}(\rho_1)=
h_{S} \sin[q(\rho_{1})]
\end{equation}
\begin{equation}
\zeta_{L}(\mbox{\boldmath $\rho$})\equiv
\zeta_{L}(\rho_1)=
h_{L} \sin[q \rho_{1}] + \ell
\end{equation}
Now the expression (A2) can be averaged over the 
\(\mbox{\boldmath $\rho$}\)-coordinates,
thus yielding the mean interaction
\(\overline{P}(\ell,[\zeta_L]) \equiv  
\langle P(\mbox{\boldmath $\rho$},[\zeta_L]) 
\rangle_{\mbox{\boldmath $\rho$}} \),
which for sinusoidal interfacial profiles reads
\begin{eqnarray}
\overline{P}(\ell,h_L) &=&
P^{0}(\ell) +
\frac{1}{4} \int \mbox{d}^{2}\mbox{\boldmath $\rho$}
\left(h_{S}^{2} + h_{L}^{2} -2h_{S}h_{L}\cos[q \rho_{1}]\right) 
w(\mbox{\boldmath $\rho$},\ell) +\nonumber \\
&&\frac{1}{64}\left.
\int \mbox{d}^{2}\mbox{\boldmath $\rho$}
\left(h_{S}^{4} + h_{L}^{4}+4h_{S}^{2} h_{L}^{2}-4h_{S}h_L(h_{S}^{2} + h_{L}^{2})
\cos[q \rho_{1}]+2h_{S}^{2} h_{L}^{2}\cos[2q \rho_{1}]\right) 
\frac{\partial^2}{\partial z^2}
w(\mbox{\boldmath $\rho$},z)\right|_{z=\ell}
\end{eqnarray}
and is an expansion up to fourth order
in the  interface modulation amplitudes
$h_S$ and $h_L$.

To further proceed, it is appropriate to specify the molecular interaction
\(w(\mbox{\boldmath $\rho$},z)\). In all what follows, an  inverse
power law defined by
\begin{equation}
w(\mbox{\boldmath $\rho$},z) \equiv A
\left( \rho^2 +z^2\right)^{-m/2}
\end{equation}
will be used.
Accordingly, non-retarded van der Waals interactions correspond
to $m=6$ with $A$ being the Hamaker constant.
The interaction between planar surfaces is given by
\begin{equation}
P^{0}(\ell)=
\frac{2 \pi A }{(m-2)(m-3)(m-4)} \;\;\; \ell\;^{4-m}
\end{equation}
In addition, the following relations involving derivatives of 
$P^{0}(\ell)$ turn out to be useful:
\begin{equation}
P^{(II)}(\ell)=\frac{2\pi A}{m-2}\ell\;^{2-m}=
A \int \mbox{d}^{2}\mbox{\boldmath $\rho$}\;\;
(\rho^2+\ell^2)^{-m/2}
\end{equation}
\begin{equation}
P^{(IV)}(\ell)=2\pi A (m-1) \;\; \; \ell^{-m}
\end{equation}
\begin{equation}
P^{(-II)}(\ell)=\frac{2\pi A }{(m-2)(m-3)(m-4)(m-5)(m-6)}\;\;\;
\ell\;^{6-m}
\end{equation}
where $\mbox{d}^2 P^{(-II)}(\ell) / \mbox{d} \ell^2= P^{0}(\ell)$.
Note that $ P^{(-II)}(\ell) $ is not defined for van der Waals 
interactions with $m=6$; this important case will be considered
separately.

The terms in (A6) involving a cosine  can now be evaluated analytically;
for the first term, and using the interaction defined in
(A7), we obtain
\begin{eqnarray}
A \int_{-\infty}^{\infty} \mbox{d} \rho_{1} 
\int_{-\infty}^{\infty} \mbox{d} \rho_{2}
\frac{\cos[q \rho_{1}]}{
\left(\rho_{1}^{2}+ \rho_{2}^{2}+\ell^{2}\right)^{m/2}}
&=&\frac{A \sqrt{\pi}\Gamma(\frac{m-1}{2})}{\Gamma(\frac{m}{2})}
\int_{-\infty}^{\infty} \mbox{d} \rho_{1}
\frac{\cos[q \rho_{1}]}{
\left(\rho_{1}^{2}+\ell^{2}\right)^{(m-1)/2}} \nonumber \\
&=&\frac{2\pi A}{\Gamma(\frac{m}{2})}
K_{m/2-1}(q\ell) (q \ell)^{m/2-1} \ell^{2-m} 2^{1-m/2}
\end{eqnarray}
In the last equation, 
$K_{\nu}(z)$ denotes the Modified Bessel Function in standard notation.

\subsection{Small separations}

Let us first concentrate on separations $\ell$ much smaller
than the typical wavelength $q^{-1}$ of the surface corrugation,
i.e., $\ell \ll q^{-1}$. For small 
values of the argument $z$, the product 
$K_{\nu}(z)z^{\nu}$ can be expanded as a power series
\begin{equation}
K_{\nu}(z)z^{\nu}=a_{0}+a_{2}z^{2}+a_{4}z^{4}+ \cdots
\end{equation}
with $\nu=m/2-1$ and the coefficients given by
\begin{eqnarray}
a_{0}&=2^{\nu-1}\Gamma(\nu)    &\;\;\;\;\;\;\mbox{for}\;\; \nu>0  \\
a_{2}&=-2^{\nu-3}\Gamma(\nu-1) &\;\;\;\;\;\;\mbox{for}\;\; \nu>1 \\
a_{4}&=2^{\nu-6}\Gamma(\nu-2)  &\;\;\;\;\;\;\mbox{for}\;\; \nu>2
\end{eqnarray}
Note that for van der Waals forces, characterized by $\nu=2$,
the coefficient $a_4$ is given by
\begin{equation}
a_4= (3/2-2\gamma +2 \log 2 -2 \log z)/16
\end{equation}
with $\gamma$ being the Euler constant defined  by $\gamma=0.57721$.
Using this expansion and the definitions (A8-A11),
the cosine term in (A6) can be written as
\begin{equation}
A \int_{-\infty}^{\infty} \mbox{d} \rho_{1} 
\int_{-\infty}^{\infty} \mbox{d} \rho_{2}
\frac{\cos[q \rho_{1}]}{
\left(\rho_{1}^{2}+ \rho_{2}^{2}+\ell^{2}\right)^{m/2}}
= P^{(II)}(\ell)-
\frac{m-3}{2} q^2 P^{0}(\ell)+
\frac{(m-3)(m-5)}{8} q^4 P^{(-II)}(\ell)
+{\cal O}(q^6)
\end{equation}
For the case of van der Waals interactions, $m=6$, one analogously obtains
\begin{equation}
A \int_{-\infty}^{\infty} \mbox{d} \rho_{1} 
\int_{-\infty}^{\infty} \mbox{d} \rho_{2}
\frac{\cos[q \rho_{1}]}{
\left(\rho_{1}^{2}+ \rho_{2}^{2}+\ell^{2}\right)^{3}}
= \frac{2 \pi}{4} \left(\frac{1}{\ell^4} -\frac{q^2}{4 \ell^2}
+\frac{3 - 4 \gamma+4 \ln 2 -4 \ln(q \ell)}{64} q^4 \right)
+{\cal O}(q^6)
\end{equation}
which includes a logarithmic singularity for small separations.

For the other term in (A6) involving 
a cosine one can interchange the differentiation
and integration (for $m>6$); the additional integral
needed involves $\cos[2 q \rho_1]$ and is given by
\begin{equation}
A \int_{-\infty}^{\infty} \mbox{d} \rho_{1} 
\int_{-\infty}^{\infty} \mbox{d} \rho_{2}
\frac{\cos[2 q \rho_{1}]}{
\left(\rho_{1}^{2}+ \rho_{2}^{2}+\ell^{2}\right)^{m/2}}
= P^{(II)}(\ell)-
2(m-3)q^2 P^{0}(\ell)+
2(m-3)(m-5) q^4 P^{(-II)}(\ell)
\end{equation}
Using the formulas (A18), (A20), and the definitions (A8)-(A11),
one obtains the following expansion  for (A6)
\begin{eqnarray}
\overline{P}(\ell,h_L)&=&
P^{0}(\ell)\left[1 + \frac{m-3}{4} h_{S}h_{L}q^2
-\frac{(m-3)(m-5)}{128}h_{S}h_{L} (h_{S}-h_{L})^2 q^4
+\frac{3(m-3)(m-5)}{64}h_{S}^2h_{L}^2q^4 \right] \nonumber\\
&& + P^{(II)}(\ell)\left[ \frac{1}{4} (h_{S}-h_{L})^2+
\frac{m-3}{32}h_{S}h_{L} (h_{S}-h_{L})^2 q^2 \right]+
P^{(IV)}(\ell)\frac{1}{64}(h_{S}-h_{L})^4 -
\delta \overline{P}(\ell,h_L)
+{\cal O}(h^6,q^6)
\end{eqnarray}
which is valid for $q \ell \ll 1$ and $m>4$ (for $m=4$
additional logarithmic singularities appear in terms proportional
to $q^2$).
The terms up to ${\cal O}(h^2, q^2)$ have
been obtained previously in the context of the
dynamics of thin wetting layers\cite{Harden}.
The correction term $\delta \overline{P}$
corresponds to a curvature contribution
and is given up to ${\cal O} (h^4, q^4)$ by
\begin{equation}
\delta \overline{P}(\ell,h_L)=
\frac{(m-3)(m-5)}{16} P^{(-II)}(\ell)   h_{S}h_{L}q^4
\end{equation}
This curvature contribution has additional non-analytic
terms for the case of non-retarded van der Waals
interactions, $m=6$; for this case, one obtains
\begin{equation}
\delta \overline{P}(\ell,h_L)=
\frac{\pi}{256}
\frac{3 - 4 \gamma+4 \ln 2 -4 \ln(q \ell)}{64} h_S h_L q^4
\end{equation}
The same singularity $\sim q^4 \ln q$ has been found for a free
interface in the presence of 
van der Waals forces.\cite{Dietrich2}

\subsection{Large separations}

In the other limit, for $q \ell \gg 1$,
the product $K_{\nu}(z)z^{\nu}$ is given by 
\begin{equation}
K_{\nu}(z)z^{\nu}=\sqrt{\pi/2} \;\; z^{\nu-1/2} 
\mbox{e}^{-z} 
(1+(4\nu^2-1)/8z + \cdots)
\end{equation}
In this case, the expression (A12) is asymptotically given by
\begin{equation}
A \int_{-\infty}^{\infty} \mbox{d} \rho_{1}
\int_{-\infty}^{\infty} \mbox{d} \rho_{2}
\frac{\cos[q \rho_{1}]}{
\left(\rho_{1}^{2}+ \rho_{2}^{2}+\ell^{2}\right)^{m/2}}
= \frac{2^{3/2-m/2}  \pi^{3/2} A }{\Gamma(\frac{m}{2})}
q^{m/2-3/2} \ell^{1/2-m/2} \mbox{e}^{-q \ell}
\end{equation}
From equation (A6)
one immediately obtains
\begin{equation}
\overline{P}(\ell,h_L)=
P^{0}(\ell)+\frac{1}{4} P^{(II)}(\ell) (h_{S}^2+h_L^2) 
 +\frac{1}{64} P^{(IV)}(\ell)(h_S^4 + h_L^4 + 4 h_S^2 h_L^2)
+{\cal O}(\mbox{e}^{-q \ell})
\end{equation}
which is valid for $q \ell \gg 1$.

\subsection{Closed-form expressions for the effective
potential}

In the following, formulas are presented, which 
express the series for the effective interaction
between the two rough interfaces, (A21) and (A26),
in terms of the interaction $P^0(\ell)$ between 
two flat interfaces.

For the case $q \ell \ll 1$, this expression is given by
\begin{equation}
P(\rho_1,\ell,h_L)=
(1+h_S h_L q^2 \cos^2[q \rho_1])^{1/2} P^0 \left(
\frac{\ell+(h_S-h_L)\sin[q \rho_1]}{
(1+h_S h_L q^2 \cos^2[q \rho_1])^{1/2}} \right) - 
\delta P(\rho_1,\ell,h_L)
\end{equation}
For the case $q \ell \gg 1$, the corresponding expression is given by
\begin{equation}
P(\rho_1,\ell,h_L)=
\langle  P^0 \left( \ell + h_L \sin[q \rho_1] + 
h_S \sin[q \tau_1]\right)
\rangle_{\tau_1}
+{\cal O}(\mbox{e}^{-q \ell})
\end{equation}
where $\tau_1$ is the local lateral coordinate on the 
liquid interface and is averaged over, leaving only
the dependence on the coordinate $\rho_1$ in the 
substrate interface.
Expressions (A27) and (A28)  depend explicitly on the spatial coordinate 
$\rho_1$; that they indeed reproduce term by term  the 
series  (A21) and (A26) can be checked by expansion and 
averaging over $\rho_1$.
The validity of the closed form expression is thus proven
for power laws with arbitrary $m$; we were not able to extend
this proof to interactions which include a cut-off at small
separations. However, it is likely that the formulae (A27) and (A28) 
are also accurate for potentials $P^0(\ell)$ which do not
diverge as $\ell \rightarrow 0$. This is supported by the fact that
for $\ell=0$ the formula (A27) exactly describes the surface-like
energy contributions.

\begin{figure}
\caption{
Schematic view of a  flat substrate:
A liquid layer of thickness $\ell$ intrudes between the
inert substrate and the top phase, which can be either
the vapor or the solid in chemical equilibrium with the liquid,
corresponding to wetting or interfacial premelting, respectively.
The liquid-substrate and liquid-top phase interfacial energies 
are denoted by $\gamma_{SL}$ and $\gamma$, respectively.}
\end{figure}

\begin{figure}
\caption{
Liquid film on a rough substrate:
For thin mean film thickness $\ell$,
defined by the averaged local separation between
the two interfaces, the liquid interface follows the
substrate corrugations.}
\end{figure}

\begin{figure}
\caption{ 
Simplified geometry in the single $q$-mode
approximation, with the substrate surface
parameterized by 
$\zeta_S(\rho_1)$, and the liquid interface
parameterized by $\zeta_L(\rho_1)$,
shown along the direction parallel to the wavevector
of the sinusoidal profile.
The mean separation $\ell$ corresponds to the distance between the
mean positions of the interfaces, denoted by  broken lines.}
\end{figure}

\begin{figure}
\caption{ 
Minimal mean film thickness $\ell$ 
for which a flat liquid interface is still possible.
The liquid interface touches the substrate surface at
isolated points, and the film thickness $\ell$ corresponds
to the characteristic  corrugation amplitude $h_S$.}
\end{figure}

\begin{figure}
\caption{
Liquid film with a thickness  $\ell'$, for
which the liquid interface with a corrugation amplitude
$h_L^*$ as given by (35) just touches the substrate.
Since the two interfaces cannot cross,
in the single $q$-mode approximation 
the amplitude $h_L$ increases
for smaller values of $\ell$
until one finally obtains  $h_L=h_S$ in the limit $\ell \rightarrow 0$.
In a more realistic model, one might obtain a ruptured interface
for small layer thicknesses, corresponding to separate droplets. }
 \end{figure}

\begin{figure}
\caption{ 
Plot of the lower bound for wetting potentials,
$-P^0_{\rm low}/S$, as a function
of the rescaled film thickness $\ell/ h_S$,
for $-S/\gamma =1/2$ and  $S_{\rm eff}=0$.
In the interval $0<\ell < 2 \ell'$,
$P^0_{\rm low}(\ell)$ is a monotonically increasing function.
In the interval  $ 2 \ell'<\ell < \ell_{\rm max}$,
$P^0_{\rm low}(\ell)$ is constant. For $\ell>\ell_{\rm max}$ 
one finds $P^0_{\rm low}(\ell)=0$. The value $\ell^*$
is  defined by $P^0_{\rm low}(\ell^*)=0$.}
\end{figure}
 
\begin{figure}
\caption{ 
Plot of the lower bound for wetting potentials,
$-P^0_{\rm low}/S$, as a function
of the rescaled film thickness $\ell/ h_S$,
in the limit $\gamma \gg -S$, for $-S/\gamma =1/2$,
and for $-S/\gamma =2/3 $ (from bottom to top).
The curves shown correspond to the special
case $S_{\rm eff}=0$, i.e., the relative area of
the substrate is just sufficient  to make the
free energy of the dry state ($\ell=0$)  higher than the
completely wet state ($\ell=\infty$).
The area ratios are given by $\alpha_S=1$,
$2$, and $3$, from bottom to top.}
\end{figure}

\begin{figure}
\caption{ 
Plot of the lower bound for wetting potentials,
$-P^0_{\rm low}/S$, as a function
of the rescaled film thickness $\ell/ h_S$,
 for the fixed value $-S/\gamma =1/2$ and for $S_{\rm eff}=0$,
 $S_{\rm eff}=-S$, and for $S_{\rm eff}=-2S$
 (from bottom to top for the right portion of the plots).
The relative area increase of the rough substrate is  given by $\alpha_S=2$,
$3$, and $4$, from bottom to top.}
\end{figure}


\begin{references}
\bibitem{foot1}
Eq. (2) corresponds to the so-called sharp-kink approximation,
where the density is constant in the two
coexisting phases and changes abruptly at the interface.
The effect of continuously varying density profiles including 
pronounced density oscillations which are expected close to 
the wall, as well as of fluctuations of the interfacial configurations  
might play a role at finite separations which are considered
in this paper.\cite{Fisher}
The present theory thus serves as a starting point
for a more complete investigation of these additional complications.
\bibitem{foot3}
In the case of volatile liquids, i.e., liquids which are in equilibrium
with either the vapor or solid phase, the true interfacial
tension between the substrate and the top phase
 is never larger than the sum of the 
two tensions $\gamma$ and $\gamma_{SL}$ (Antonov's rule), since a liquid
layer would always form  spontaneously; in fact, the interfacial tension
between the third phase and the bare substrate cannot even be measured if
the spreading coefficient $S$, as defined by (4), is positive.
Hence, the measurable spreading coefficient is always negative.
\bibitem{foot4}
Experimentally, glass substrates can be roughened by 
exposing them to HF/Nitric acid, where the surface height fluctuations
are determined by the etching  time\cite{Beaglehole2}.
\bibitem{foot5}
Here one neglects curvature effects
which to leading order are proportional to $(\nabla^2 
\zeta_S(\mbox{\protect \boldmath \protect $\rho$}))^2$.
This approximation is permitted if the substrate corrugation
is dominated by small wave numbers, see \cite{estimate}. 
The curvature correction  
for various molecular interactions is calculated in the Appendix.
\bibitem{estimate}
In the experiment demonstrating roughness-induced interfacial
premelting of ice\cite{Beaglehole2}, the characteristic height
fluctuations have amplitudes and wavenumbers in the range
of $h_S \approx 2 nm$ and $q \approx 0.01 nm^{-1}$, leading
to $h_S q \approx 0.02$.
\end{references}
\end{document}